\documentclass[showpacs,pra,twocolumn,aps]{revtex4}

 \usepackage{bm}
 \usepackage{amsmath}
 \usepackage{amssymb}
 \usepackage{latexsym} 
 \usepackage{amsfonts} 
 \usepackage{epsfig}
 \usepackage{psfrag} 

 \newcommand{\ket}[1]{\left|#1\right>}

 \newcommand{\nn}{\nonumber\\} 
  
 \newcommand{\f}[1]{\mbox{\boldmath$#1$}}
 \newcommand{\fk}[1]{\mbox{\boldmath$\scriptstyle#1$}}
 \newcommand{\vau}{\mbox{\boldmath$v$}}
 \newcommand{\na}{\mbox{\boldmath$\nabla$}}
 \newcommand{\bea}{\begin{eqnarray}}
 \newcommand{\ea}{\end{eqnarray}}
 \newcommand{\eea}{\end{eqnarray}}
 \newcommand{\ord}{{\cal O}}

\begin{document}

\title{Quantum simulation of cosmic inflation 
in two-component Bose-Einstein condensates}
 \author{Uwe R.~Fischer$^{1}$ and Ralf Sch\"utzhold$^2$}
 \affiliation{$^1$Eberhard-Karls-Universit\"at T\"ubingen, 
 Institut f\"ur Theoretische Physik, \\  
 Auf der Morgenstelle 14, D-72076 T\"ubingen, Germany\\
 $^2$Institut f\"ur Theoretische Physik, 
 Technische Universit\"at Dresden, 
 D-01062 Dresden, Germany}

\begin{abstract} 
Generalizing the one-component case, we demonstrate that the
propagation of sound waves in two-component Bose-Einstein condensates
can also be described in terms of effective sonic geometries under
appropriate conditions. 
In comparison with the one-component case, the two-component 
setup offers more flexibility and several advantages.
In view of these advantages, we propose an experiment in which
the evolution of the inflaton field, and thereby the generation of 
density fluctuations in the very early stages of our universe
during inflation, can be simulated, realizing a 
{\em quantum simulation via analogue gravity models}.
\end{abstract}
  
\maketitle
 
%%%%%%%%%%%%%%%%%%%%%%%%%%%%%%%%%%%%%%%%%%%%%%%%%%%%%%%%%%%%%%%%%%%%%%%%%%%%%%%
%%%%%%%%%%%%%%%%%%%%%%%%%%%%%%%%%%%%%%%%%%%%%%%%%%%%%%%%%%%%%%%%%%%%%%%%%%%%%%%
\section{Introduction}\label{Introduction}
%%%%%%%%%%%%%%%%%%%%%%%%%%%%%%%%%%%%%%%%%%%%%%%%%%%%%%%%%%%%%%%%%%%%%%%%%%%%%%%
%%%%%%%%%%%%%%%%%%%%%%%%%%%%%%%%%%%%%%%%%%%%%%%%%%%%%%%%%%%%%%%%%%%%%%%%%%%%%%%

Within our present standard model of cosmology, basically all
inhomogeneities -- including the seeds for the formation of structures
such as our galaxy -- originate from quantum fluctuations of a single
scalar field, the {\em inflaton}  \cite{Linde,Lidsey}. 
This (postulated) field drives inflation, which is a stage of very
rapid expansion in the earliest evolutionary phase of our universe
\cite{inflation}.  
Tracing the inflaton fluctuations back in time and thereby undoing the
redshift induced by the cosmic expansion, the anisotropies of the 
cosmic microwave background we observe today correspond to 
extremely short wavelengths during inflation.
As a result, the fluctuations of the cosmic microwave background 
probe ultra-high (e.g., Planckian) energy scales -- which are
experimentally inaccessible with the present-day and near-future 
available technology of, for example, particle accelerators. 
At such ultra-high energies, quantum effects of
gravity are expected to become important -- but the underlying
physical theory for the description of these effects is not 
known yet.
Consequently, high-precision measurements of the cosmic microwave
background might give us some insight into new physics beyond
well-established theories (in particular, beyond the standard model 
of particle physics). 

In order to detect signatures of the new physics in 
the anisotropies of the cosmic microwave background, one has 
to investigate which kind of higher-order corrections and 
correlations could potentially be induced by deviations from 
the known laws of physics occurring at ultra-high energies.
One way to achieve this aim is to consider analogous systems, based on 
laboratory physics, which reproduce major features of the inflaton
field, and which can therefore be used to simulate the generation 
of (quantum) fluctuations and further interesting effects --
theoretically as well as experimentally. 
This line of approach has come to be known under the term 
{\em analogue gravity/cosmology}, see, e.g., \cite{GrishaBook}.
The consideration of these analogues leads to a better understanding
of the system to be simulated, in particular, 
regarding the possible impact of high-energy
degrees of freedom.
As cosmology is essentially a purely observational branch of science, because 
obviously we cannot do experiments on the real system, i.e., the universe, 
the analogues also allow an experimental verification of so far only 
theoretically predicted effects.
Basically the same motivations underlie the idea of 
quantum black hole analogues
(''dumb holes'' \cite{unruh,MattPRL,AG-BEC}), since the Hawking radiation 
of ``real'' gravity black holes 
has its origin in (trans-)Planckian modes: The quanta emitted
necessarily come from regions very close to the black hole horizon, 
and experience a large redshift when finally detected far away from the 
black hole \cite{unruh,BHdispers}.

A scalar field (such as the inflaton) within the curved space-time
structure of an expanding universe can be simulated by propagating
sound waves in single-component Bose-Einstein  condensates (BEC)
\cite{remark};  
previous considerations on such effective acoustic
geometries in single-component BECs can be found 
in Refs. \cite{AG-BEC,BLVBEC,uwe+petr,0303063,BLVPRA}.
The advantage of BECs lies in the fact that the 
corresponding parameters (such as local density and speed of sound)
can be controlled with atomic precision experimentally, and that the
underlying  physics is well-understood on all energy scales
in particular in the dilute case, for which the Gross-Pitaevski\v\i\/
equation provides a rather accurate description of the
order parameter dynamics.
There are basically two possibilities for simulating an expanding 
universe within a BEC: changing the interparticle coupling or 
expanding the condensate (or a combination of both) 
\cite{uwe+petr,0303063,BLVPRA}.
However, both methods come with problems.
Firstly, a controlled expansion of the BEC cloud requires a specific 
time-dependent trap; and 
furthermore, the density of the cloud rapidly decreases during the expansion, 
leaving only a short time to do the experiment.
Secondly, in order to change the inter-particle coupling drastically via a 
Feshbach resonance, by rapidly 
sweeping a time-dependent external magnetic field, 
one has to go very close to the resonance; one then encounters
the problem that the coupling constant effectively acquires an 
imaginary part due to molecule formation in three-body 
recombination processes \cite{Piotr,Weber}, which spoils the 
desired effect.
Thirdly, a time-dependent interparticle coupling would also induce a 
variation of the density and the size of the BEC cloud -- unless the 
trap is changed accordingly, which again is difficult (see first point).
Finally, one has to be able to measure the generated fluctuations in order to 
simulate the inflaton field experimentally.

In the present article, we investigate whether it is possible to overcome 
some of these problems in two-component Bose-Einstein condensates,
which are readily experimentally available; the various components 
can be realized by trapping different hyperfine ground states of the same atom 
\cite{HoShenoy,Timmermans,Hall,Miesner}.
We start by describing in section \ref{Equations of motion} how the collective 
equations of motion for small fluctuations (i.e., sound waves) in  the 
two-component gas are obtained.
In section \ref{Effective geometry}, we show that it is possible to map these
equations of motion onto two effective sonic metrics (bi-metricity)
under certain conditions.
The analogue of the Planck scale, where the concept of the effective geometry
breaks down, is discussed in section \ref{Dispersion}.
Section \ref{Inflaton} is devoted to the simulation of the de Sitter 
geometry -- i.e., to constructing 
an analogue for the inflaton field during inflation.
We discuss possible experimental realizations and measurement prescriptions
to realize the desired inflation quantum simulation  
in section \ref{Fluctuations}.
The advantages and drawbacks of the described method and further aspects
are summarized in section \ref{Conclusions}.

We note that the inflaton has been mentioned in the context of Bose-Einstein 
condensates previously \cite{CalzettaHu}. 
However, the discussion there has been rather qualitative. 
In particular, the inflaton mode has not been related there to any effective 
space-time metric of cosmological character (for example the 
de Sitter metric), in which 
its propagation ought to take place. 
The latter is necessary to appropriately 
describe the freezing process of the quantum 
fluctuations and the related concept of a horizon, both of which 
we shall investigate for de Sitter space-time in what follows. 

%%%%%%%%%%%%%%%%%%%%%%%%%%%%%%%%%%%%%%%%%%%%%%%%%%%%%%%%%%%%%%%%%%%%%%%%%%%%%%%
%%%%%%%%%%%%%%%%%%%%%%%%%%%%%%%%%%%%%%%%%%%%%%%%%%%%%%%%%%%%%%%%%%%%%%%%%%%%%%%
\section{Equations of motion}\label{Equations of motion}
%%%%%%%%%%%%%%%%%%%%%%%%%%%%%%%%%%%%%%%%%%%%%%%%%%%%%%%%%%%%%%%%%%%%%%%%%%%%%%%
%%%%%%%%%%%%%%%%%%%%%%%%%%%%%%%%%%%%%%%%%%%%%%%%%%%%%%%%%%%%%%%%%%%%%%%%%%%%%%%

In terms of the Madelung representation for the order parameter 
components $\psi_a(\f{r},t)$ 
\bea
\psi_a(\f{r},t)=\sqrt{\varrho_a(\f{r},t)}\exp\{iS_a(\f{r},t)\}
\,,
\ea
with the density $\varrho_a(\f{r},t)$ and the phase (eikonal)
$S_a(\f{r},t)$,  
the Lagrangian density of a dilute two-component Bose-Einstein condensate 
reads (we put $\hbar=1$) \cite{HoShenoy}
\bea
{\cal L}
& = & 
-\sum_a
\left(
\varrho_a\partial_t S_a
+
\frac{\varrho_a}{2m_a}(\na S_a)^2
+
\frac{(\na\sqrt{\varrho_a})^2}{2m_a}
+
V_a\varrho_a 
\right)
\nonumber\\
& & -\frac12\sum_{a,b}g_{ab}\varrho_a\varrho_b
\,.
\ea
Here, the two masses of the atoms are $m_a$ ($a=1,2$), 
the one-particle trapping potentials (which are
generally different) are given by $V_a(\f{r},t)$, 
and the (symmetric) two-particle interaction coupling matrix 
is denoted $g_{ab}$.  
 
Linearizing around a given, stationary background solution
 $\varrho_a^0(\f{r})$ and $S_a^0(\f{r})$ with $\vau^0_a=\na S_a^0/m_a$
 and neglecting the quantum pressure terms $\propto
 (\na\sqrt{\varrho_a})^2$ -- which amounts to the local density   
 (Thomas-Fermi) approximation in one-component Bose-Einstein
 condensates -- leads to the second-order effective action 
 \bea
 {\cal L}_{\rm eff}^{(2)}
 & = & 
 -\sum_a 
\left[
\delta\varrho_a \partial_t \delta S_a
+ 
\frac{\varrho_a^0}{2m_a}(\na\delta S_a)^2
+
\delta\varrho_a\vau^0_a\cdot\na\delta S_a \right]
\nn 
&&
-\sum_{a,b}\frac12\,g_{ab}\delta\varrho_a\delta\varrho_b 
\,.
\ea
Varying the above action with respect to $\delta\varrho_a$ yields two
Bernoulli-type equations for the fluctuations
 \bea
\label{Bernoulli}
 D_a  \delta S_a
 + \sum_b g_{ab}\delta\varrho_b=0
\,, 
\ea
 where the co-moving derivative is defined to be 
 \bea
 D_a\delta S_a=\partial_t\delta S_a+\vau^0_a\cdot\na\delta S_a
\,.
 \ea
 Using the above equation to eliminate $\delta\varrho_a$
 from ${\cal L}_{\rm eff}^{(2)}$, we obtain a phases-only 
 effective Lagrangian of the form 
 \bea
\label{phases-only}
 {\cal L}_{\rm eff}^S
 = \sum_{ab}
 \frac12(D_a\delta S_a)g_{ab}^{-1}(D_b\delta S_b)
 -\sum_a\frac{\varrho_a^0}{2m_a}(\na\delta S_a)^2 
\,.
 \ea
The general wave equations for the phase fluctuations $\delta S_a$
then take the form
\begin{equation}
\sum_b D_a \left(g_{ab}^{-1} D_b \delta S_b \right)
-\frac{\varrho_a^0}{m_a} \na^2 \delta S_a= 0   
\,.
\end{equation}
These wave equations, in the general case, do not yet have the 
pseudo-Lorentz invariance required to obtain effective space-time 
metrics of Lorentzian signature. 

%%%%%%%%%%%%%%%%%%%%%%%%%%%%%%%%%%%%%%%%%%%%%%%%%%%%%%%%%%%%%%%%%%%%%%%%%%%%%%%
%%%%%%%%%%%%%%%%%%%%%%%%%%%%%%%%%%%%%%%%%%%%%%%%%%%%%%%%%%%%%%%%%%%%%%%%%%%%%%%
\section{Effective geometry}\label{Effective geometry}
%%%%%%%%%%%%%%%%%%%%%%%%%%%%%%%%%%%%%%%%%%%%%%%%%%%%%%%%%%%%%%%%%%%%%%%%%%%%%%%
%%%%%%%%%%%%%%%%%%%%%%%%%%%%%%%%%%%%%%%%%%%%%%%%%%%%%%%%%%%%%%%%%%%%%%%%%%%%%%%

So far, we have involved no specific assumptions about the background
densities and velocities as well as the constituent masses. 
We now come to discuss a simple case in which an effective metric
description in terms of the Painlev\'e-Gullstrand-Lema{\^\i}tre type 
\cite{PGL} is viable. 

As demonstrated in Ref.~\cite{BLV2002}, the introduction of effective 
geometries for multiple interacting fields is more involved than the 
single-field case -- where rather general assumptions ensure the 
existence of an effective metric for the propagation of perturbations.
In the case of multiple interacting fields, there are the following
three main possibilities:
{\bf a)} Owing to a lack of symmetry one cannot introduce a metric at all 
(``pre-geometry'');
{\bf b)} the perturbations effectively decouple and can be described by 
multiple metrics; or
{\bf c)} all the metrics coincide and there is one unique metric 
(reproducing the principle of equivalence). 
There is also the (fourth) possibility that the propagation of 
perturbations is not equivalent to scalar (i.e., spin-zero) fields 
in curved space-times, but to fields with higher, non-zero spin instead
(e.g., Dirac \cite{slow-light} or vector \cite{dbha} fields).

Phonons in arbitrary two-component condensates correspond to case 
{\bf a)} in general -- 
even though one might diagonalize the dispersion relation, 
the full equations of motion do not allow the
introduction of an effective geometry (in the most general situation). 
In order to arrive at an effective metric, certain requirements on the
background solution are necessary.
Let us assume that the parameters of the background solution satisfy 
the following conditions
\bea
\vau_1^0 
& = & 
\vau_2^0 \;\equiv\; \vau 
\quad \leadsto \quad  
D_1 = D_2 \equiv D
\nn
\frac{\varrho_1^0}{m_1} 
& = & 
\frac{\varrho_2^0}{m_2} \;\equiv\; \frac{\varrho_0}{m}
\nn
{\rm eigenvectors}\,(g_{ab}) & = & {\rm const.} 
\label{Req1}
\ea
The diagonalization of the (real and symmetric) coupling matrix
$g_{ab}$ leads to eigenvalues $g_{\pm}$, given by
\bea
g_{\pm} = \frac{g_{11}+g_{22}}{2} \pm 
\sqrt{\left(\frac{g_{11}-g_{22}}{2}\right)^2 + g_{12}^2}
\,.
\ea
Note that -- in contrast to the eigenvectors of the matrix $g_{ab}$ --
its eigenvalues $g_{\pm}$ are not required to be constant.
By virtue of the assumptions (\ref{Req1}), the Lagrangian in 
Eq.~(\ref{phases-only}) can be diagonalized
\bea
\label{effLag}
 {\cal L}_{\rm eff}^S
 & = & 
\frac12\,g_{\pm}^{-1}(D\phi_\pm)^2-\frac{\varrho_0}{2m}(\na\phi_\pm)^2
 \nonumber\\
 & \equiv & 
 \frac12 \sqrt{-{\sf g}_\pm}\, {\sf g}^{\mu\nu}_\pm  \partial_\mu \phi_\pm 
 \partial_\nu
 \phi_\pm
\,,
\ea
where $\phi_\pm$ denote the projections of the phase fluctuations
$\delta S_a$ onto the (constant) eigenvectors of $g_{ab}$ and a 
summation convention over indices $\mu,\nu$ and $\pm$ is implied.
In this (highly symmetric) case, we therefore obtain two 
{\em independently propagating}, i.e., decoupled modes $\phi_\pm$,
which feel effective space-time metrics of the conventional
(covariant) Painlev\'e-Gullstrand-Lema{\^\i}tre form \cite{visser} 
\begin{equation}
 {\sf g}^\pm_{\mu\nu}= \frac{\varrho_0}{c_\pm} 
 \left( \begin{array}{cr} c_\pm^2-{\bm v}^2 \, &\, {\bm v}  
 \\ {\bm v}\, & -{\bm 1} 
 \end{array} \right)
\,,
\end{equation} 
where the two sound velocities are  $c_{\pm} = \sqrt{g_\pm\varrho_0/m}$
and $\bm 1$ represents the unit matrix. 
As long as these two sound velocities do not coincide $c_+ \neq c_-$,
the system under consideration corresponds to the bi-metric case 
{\bf b)} discussed at the beginning of this Section.

The assumption in (\ref{Req1}) that the eigenvectors of $g_{ab}$ be
constant can be satisfied, if $g_{ab}$ itself is constant, or if it is
sufficiently symmetric. 
We shall assume $g_{11} = g_{22}$ in the following, because it allows
both for a bi-metric approach and an implementation of time-dependent 
$g_\pm=g_\pm(t)$.  
In this situation of $g_{11} = g_{22} \equiv g_{\rm diag}$, the
eigenvalues are simply given by 
$g_{\pm} = g_{\rm diag} \pm g_{\rm off}$, where
$g_{12} = g_{21} \equiv g_{\rm off}$, and the eigenvectors read 
\bea
\label{eigenvectors}
\phi_\pm = \frac{\delta S_1 \pm \delta S_2}{\sqrt{2}}
\,.
\ea

If, in addition, $g_{\rm diag} \approx g_{\rm off}$,  
which can be fulfilled to a high degree of accuracy 
(on the level of three percent) between different 
hyperfine species in $^{87}\!$Rb (Ref.\,\cite{Hall}) as well as in 
$^{23}\!$Na (Ref.\,\cite{Miesner}), we have one ``hard'' density mode  
$\phi_+ $, and one  ``soft'' spin mode $\phi_-$. 
This separation of energy scales occurs close to the point of spatial 
phase separation of the two components, due to the  increased 
interspecies repulsion $g_{\rm off}$ \cite{HoShenoy,Timmermans}.    

%%%%%%%%%%%%%%%%%%%%%%%%%%%%%%%%%%%%%%%%%%%%%%%%%%%%%%%%%%%%%%%%%%%%%%%%%%%%%%%
%%%%%%%%%%%%%%%%%%%%%%%%%%%%%%%%%%%%%%%%%%%%%%%%%%%%%%%%%%%%%%%%%%%%%%%%%%%%%%%
\section{Dispersion relation and the Planck scale}\label{Dispersion}
%%%%%%%%%%%%%%%%%%%%%%%%%%%%%%%%%%%%%%%%%%%%%%%%%%%%%%%%%%%%%%%%%%%%%%%%%%%%%%%
%%%%%%%%%%%%%%%%%%%%%%%%%%%%%%%%%%%%%%%%%%%%%%%%%%%%%%%%%%%%%%%%%%%%%%%%%%%%%%%

So far we discussed the effective geometry for low-energy excitations 
(sound waves).
The full dispersion relation, without the Thomas-Fermi approximation
(i.e., not neglecting the quantum pressure terms) can be obtained via 
the JWKB approximation, which amounts to the geometrical optics limit
of quasiparticle propagation. 

Combining the (linearized) equation of continuity
\bea
i\left(\omega+\vau^0_a\cdot\f{k}\right)\delta\varrho_a
=
\frac{\varrho_a^0}{m_a}\,\f{k}^2\delta S_a
\,,
\ea
and the Bernoulli-type equation (\ref{Bernoulli}) augmented 
with the quantum pressure term on the
right-hand side,  
\bea
i\left(\omega+\vau^0_a\cdot\f{k}\right)\delta S_a
+
\sum_b g_{ab}\delta\varrho_b
=
-\frac{\f{k}^2}{4m_a\varrho_a^0 }\,\delta\varrho_a
\,,
\ea
gives us two Bogoliubov dispersion relations of the usual type.
Using the assumptions (\ref{Req1}), we obtain  
\bea
\left(
\frac{m}{\varrho_0}\left(\omega_\pm+\vau \cdot\f{k}\right)^2
+
\frac{\f{k}^4}{4m\varrho_0 }
\right)
=
g_\pm\f{k}^2
\,.
\ea
The deviation from the linear (rest frame) dispersion 
$\omega^2\propto\f{k}^2$ occurs at the two healing lengths 
\bea
\xi_\pm^2=\frac{1}{4m \varrho_0 g_\pm} 
\,. \label{healing}
\ea
Accordingly, the analogues of the Planck length, the two coherence lengths 
$\xi_\pm$, scale 
in the same way as the inverse spin and density mode velocities, 
\bea
\xi_\pm^2c_\pm^2=\frac{1}{4m^2} 
\,.
\ea
%
%%%%%%%%%%%%%%%%%%%%%%%%%%%%%%%%%%%%%%%%%%%%%%%%%%%%%%%%%%%%%%%%%%%%%%%%%%%%%%%
%%%%%%%%%%%%%%%%%%%%%%%%%%%%%%%%%%%%%%%%%%%%%%%%%%%%%%%%%%%%%%%%%%%%%%%%%%%%%%%
\section{De Sitter space-time and analogue inflaton}\label{Inflaton}
%%%%%%%%%%%%%%%%%%%%%%%%%%%%%%%%%%%%%%%%%%%%%%%%%%%%%%%%%%%%%%%%%%%%%%%%%%%%%%%
%%%%%%%%%%%%%%%%%%%%%%%%%%%%%%%%%%%%%%%%%%%%%%%%%%%%%%%%%%%%%%%%%%%%%%%%%%%%%%%

The extreme dependence of the sound velocity of the spin mode on the
coupling matrix $g_{ab}$ can be profitably used to simulate a rapidly
expanding universe via small temporal changes $g_{ab} = g_{ab} (t)$.
A ``spin horizon'' for the spin mode should be easier to realize
experimentally than the sound horizon of a one-component BEC, in view of
the possibility to manipulate the ``spin'' velocity such that it
closely approaches zero.

The line elements for a background at rest, $\vau=0$, read 
\bea
ds^2_\pm=\varrho_0\left(c_\pm\,dt^2-\frac{1}{c_\pm}\,d\f{r}^2\right)
\,.
\ea
From now on we focus on one particular mode, the spin mode,  
drop the subscripts $\pm$ in most of the following formulas, 
and furthermore set $m=1$ for convenience.
We suppose the background density $\varrho_0$ to remain essentially
constant during rapid variations of $g_-$ 
(which is possible for $V_a=V_b$, cf.~section \ref{Conclusions}).  

Assuming a time dependence of the propagation velocities/coupling 
constants
\bea
\label{c(t)}
c=\frac{c_0}{H^2t^2} \quad 
\Longleftrightarrow \quad g=\frac{g_0}{H^4t^{4}} 
\,,
\ea 
with $H$ being the condensed-matter 
analogue of the Hubble parameter in cosmology,
we obtain the de Sitter metric 
\bea
ds^2=\varrho_0c_0\left(d\tau^2-
\left[\frac{e^{H\tau}}{c_0}\right]^2
d\f{r}^2\right)
\,, \label{deSitter}
\ea
with a transformed de Sitter time-coordinate $\tau=H^{-1}\ln(Ht)$, 
representing proper time ({not} equal to the laboratory time),  
and the prefactor $\varrho_0c_0=\rm const$.

The corresponding Klein-Fock-Gordon equation
\bea
\label{kfg}
\Box\phi=\frac{1}{\sqrt{{\sf g}_-}}\,
\partial_\mu\left({\sqrt{{\sf g}_-}}{\sf g}_-^{\mu\nu}
\partial_\nu\phi\right)=0
\,,
\ea
with ${\sf g}_-$ denoting the (negative) determinant of the metric 
${\sf g}^-_{\mu\nu}$ and ${\sf g}_-^{\mu\nu}$ its inverse, 
assumes in this case the simple form (in three spatial dimensions)
\bea
\left(
\frac{\partial^2}{\partial\tau^2}
+
3H\,\frac{\partial}{\partial\tau}
-
e^{-2H\tau}
\left[c_0\na\right]^2
\right)\phi=0
\,.
\ea
After a spatial mode expansion into plane waves (taking into account  
isotropy and homogeneity),
each mode behaves as a damped harmonic oscillator with a time-dependent
potential 
\bea
\label{damping}
\left(
\frac{\partial^2}{\partial\tau^2}
+
3H\,\frac{\partial}{\partial\tau}
+
e^{-2H\tau}
\left[c_0\f{k}\right]^2
\right)\phi_{\fk{k}}=0
\,.
\ea
The evolution induced by the above equation of motion can roughly be 
split up into three regimes: 
\begin{itemize}
\item oscillation: For very early times, we have 
$e^{-2H\tau}[c_0\f{k}]^2 \gg H^2$; hence the damping term can
be neglected and the modes oscillate almost freely.
\item horizon-crossing: At some point in time, the monotonously
decreasing term $e^{-2H\tau}$,  corresponding to the expansion of the
universe, becomes small enough and the damping term starts to play a
role.
Since the length scale $c_0/H$ corresponds to the size of the 
particle horizon
in an expanding (de Sitter) universe, this is the point where the
modes cross the horizon and hence do not oscillate freely anymore.
\item freezing: For late times, we have 
$e^{-2H\tau}[c_0\f{k}]^2 \ll H^2$, and hence the potential term
can be neglected.  
This corresponds to a strongly over-damped oscillator and thus the
modes do effectively not evolve anymore. 
\end{itemize}
The same three essential stages undergoes the 
inflaton field during the epoch of inflation in the very early universe,  
according to our (present) standard model of cosmology \cite{Lidsey}.
Since the Hubble parameter during inflation is expected to be 
significantly smaller than the Planck scale -- otherwise the
(semi-classical) notion of a (quantum) field within a curved
space-time would not apply -- the modes leave the Planck regime 
(are ``born'') nearly freely oscillating.
Later on, with increasing $\tau$, they cross the horizon and freeze.
At the end of inflation, the frozen quantum fluctuations of the
inflaton field are transferred into (classical) density fluctuations 
(decoherence of the quantum state due to interaction with other 
degrees of freedom), which are in turn supposed to be the seeds 
for structure formation in our universe represented, e.g., by our galaxy.

%%%%%%%%%%%%%%%%%%%%%%%%%%%%%%%%%%%%%%%%%%%%%%%%%%%%%%%%%%%%%%%%%%%%%%%%%%%%%%%
%%%%%%%%%%%%%%%%%%%%%%%%%%%%%%%%%%%%%%%%%%%%%%%%%%%%%%%%%%%%%%%%%%%%%%%%%%%%%%%
\section{Measurement of Fluctuations}\label{Fluctuations}
%%%%%%%%%%%%%%%%%%%%%%%%%%%%%%%%%%%%%%%%%%%%%%%%%%%%%%%%%%%%%%%%%%%%%%%%%%%%%%%
%%%%%%%%%%%%%%%%%%%%%%%%%%%%%%%%%%%%%%%%%%%%%%%%%%%%%%%%%%%%%%%%%%%%%%%%%%%%%%%

\begin{center}
\begin{figure}[hbt]
\psfrag{Phi}{\large Im$[\phi_{\bm k}(t)]$}
\psfrag{t}{\Large $\frac{H^2 t}{c_0 k}$}
\psfrag{nk}{$|\phi^0_{\bm k}|$} 
\centerline{\epsfig{file=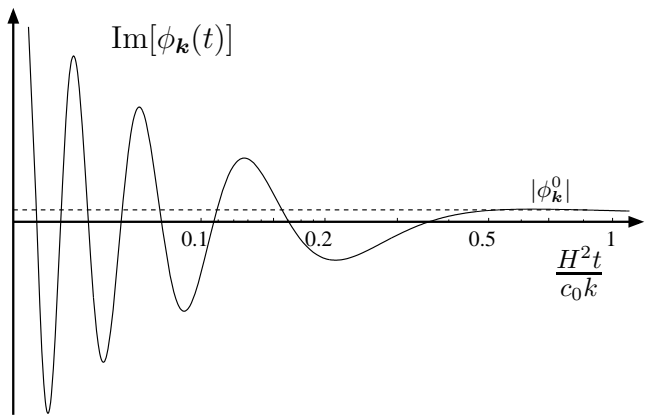,width=0.475\textwidth}}
\caption{\label{freezing} 
Freezing process for the modes $\phi_{\bm k}$ according to expression
(\ref{phieta}), with the absolute value of the frozen 
$\phi^0_{\bm k}$ given by (\ref{phasefluct}), 
displayed on a logarithmic lab time scale $t$.}
% (in arbitrary units). }
\end{figure}
\end{center}

In order to discuss the quantum state of the fluctuations $\hat\phi$, 
it is convenient to introduce yet another time-coordinate, i.e., the
conformal time $\eta=-e^{-H\tau}/H = -1/(H^2 t)$, 
in terms of which the de Sitter metric (\ref{deSitter})
can be cast into the conformally flat form
\bea
ds^2=\frac{\varrho_0c_0}{H^2\eta^2}\left(d\eta^2-
\frac{1}{c_0^2}\,d\f{r}^2\right)
\,.
\ea
Expanding the phase operator $\hat\phi$ into plane-wave solutions of the 
Klein-Fock-Gordon equation (\ref{kfg}) in conformal time 
\begin{equation}
\left(
\frac{\partial^2}{\partial\eta^2}
- \frac 2\eta \,\frac{\partial}{\partial\eta}
+ \left[c_0\f{k}\right]^2
\right)\phi_{\fk{k}}(\eta)=0 \label{conformaldamping} 
\,,
\end{equation} 
we have, for an arbitrarily chosen quantization volume $V$,
the analytical expression   
\bea
\hat\phi ({\bm r},\eta) &=& H\sqrt{\frac{g_0}{2Vc_0^3}}\sum\limits_{\fk{k}}
\frac{i- c_0 k\eta}{\sqrt{k^3}}\,e^{i\fk{k}\cdot\fk{r}-ic_0 k\eta}
\hat a_{\fk{k}}  +{\rm H.c.} \nonumber\\
\label{phieta}
\ea
We introduced creation and annihilation operators 
$\hat a^\dagger_{\fk{k}}$ and $\hat a_{\fk{k}}$, 
where the  $\hat a_{\fk{k}}$ annihilate the  
``adiabatic'' vacuum state $\ket{0}_{\rm ad}$
(see, e.g., \cite{birrell})
\bea
\hat a_{\fk{k}}\ket{0}_{\rm ad}=0
\,.
\ea
For early times (in the oscillating regime $\eta\downarrow-\infty$), 
the adiabatic quantum vacuum 
state coincides to zeroth order with the instantaneous
ground state and is therefore a natural candidate for the vacuum state.
As indicated above, after horizon crossing, the fluctuations are
frozen at late times $t\uparrow\infty\leadsto\eta\uparrow0$.
In this regime, we obtain from (\ref{phieta}) the expectation value
\bea
\langle\hat\phi_{\fk{k}}^\dagger\hat\phi_{\fk{k}}\rangle
=\frac{H^2g_0}{2Vc_0^3k^3} \equiv |\phi^0_{\bm k}|^2
\,. \label{phasefluct}
\ea
We display the evolution of a $\phi_{\bm k}$ component of
(\ref{phieta}) and the approach to the above ``frozen'' value 
in Fig.\,(\ref{freezing}).

In complete analogy to the cosmic inflaton field, the expression 
(\ref{phasefluct}) generates a scale-invariant spectrum: In view of the
$d^3k$-integration, the corresponding spatial correlation function is
(within the region of validity of the above approximations) 
invariant under re-scaling
$\langle\hat\phi_-(\f{r})\hat\phi_-(\f{r'})\rangle\simeq
\langle\hat\phi_-(\lambda\f{r})\hat\phi_-(\lambda\f{r'})\rangle$,
see, e.g., \cite{peebles}.
\vspace*{-0.5em}
%%%%%%%%%%%%%%%%%%%%%
%%%%%%%%%%%%%%%%%%%%%%%
\vspace*{-2em}

%%%%%%%%%%%%%%%%%%%%%%%%%%%%%%%%%%%%%%%%%%%%%%%%%%%%%%%%%%%%%%%%%%%%%%%%%%%%%%%
\subsection{Phase-phase correlations}
%%%%%%%%%%%%%%%%%%%%%%%%%%%%%%%%%%%%%%%%%%%%%%%%%%%%%%%%%%%%%%%%%%%%%%%%%%%%%%%

Let us investigate the impact of the frozen quantum fluctuations on the 
(spatially) Fourier-transformed two-point phase-phase correlation
function defined as
\bea
\label{phase-phase}
C_\phi(\f{k},t)
=
\int d^3r\,\exp\{i\f{k}\cdot\f{r}\}
\langle\hat\phi_-({\bm 0},t)\hat\phi_-(\f{r},t)\rangle
\,.
\ea
Inserting Eq.~(\ref{phasefluct}) we obtain (after freezing)
\bea
C_\phi(\f{k})=\frac{H^2g_0}{2c_0^3k^3}
\,.
\ea
It appears that one could generate an arbitrarily strong effect by 
increasing the Hubble parameter $H$, but this is a fallacy.
The validity of the approximations made requires that the conditions
$c_0k \gg H \gg c_{\rm final}k$ on the frequency scales hold 
(with $c_{\rm final}$ denoting the final 
value of the time-dependent speed of sound), as well as 
that the long wavelength limit $k\xi_-(t)\ll1$ is fulfilled. 
In order to estimate the maximally obtainable effect, we insert 
$H \sim c_0k$ as well as $k \sim k_{\rm final}^{\rm Planck}$ with 
$k_{\rm final}^{\rm Planck}$ denoting the inverse of the final healing
length, i.e., the final Planck scale.
The maximum effect of the dimensionless quantity $(\Delta k)^3C_\phi(\f{k})$,
i.e., the correlation within some wave-number interval $(\Delta k)^3$, 
then reads 
\bea
(\Delta k)^3C_\phi^{\rm max}(\f{k})
&\sim&
\sqrt{\frac{g_0}{g_{\rm final}}}\,
\frac{(\Delta k)^3}{\varrho_0}
\nn
&\sim&
\sqrt{\frac{g_0}{g_{\rm final}}}\,
\sqrt{\varrho_0 g_{\rm final}^3} \,
\left(\frac{\Delta k}{k_{\rm final}^{\rm Planck}}\right)^3
. \label{maxCorrphi}
\ea
Since the concept of the effective geometry only applies for
$k \ll k_{\rm final}^{\rm Planck}$, the maximum impact of the frozen 
quantum fluctuations on the two-point phase-phase correlation function
is limited by the product of the ratio 
$\sqrt{g_0/g_{\rm final}}\gg1$ and the final diluteness parameter 
$\sqrt{\varrho_0 g_{\rm final}^3}$ -- 
which is bound to be a quite small quantity; for 
the spin mode of relevance here it is of order $10^{-2}\cdots 10^{-4}$.

%%%%%%%%%%%%%%%%%%%%%%%%%%%%%%%%%%%%%%%%%%%%%%%%%%%%%%%%%%%%%%%%%%%%%%%%%%%%%%%
\subsection{Density-density correlations}
%%%%%%%%%%%%%%%%%%%%%%%%%%%%%%%%%%%%%%%%%%%%%%%%%%%%%%%%%%%%%%%%%%%%%%%%%%%%%%%

In contrast to phase-phase correlations, which can only be measured 
by time-of-flight measurements \cite{tof},  
density-density correlations can be obtained directly, 
i.e., {\em in situ}, for example via (state-selective) 
absorption (such in situ measurements are generally
more difficult to be carried out, though \cite{tof}).

The spatially Fourier-transformed two-point density-density correlation 
function is defined, in a manner analogous to its phase-phase counterpart
in (\ref{phase-phase}), to be 
\bea
\label{density-density}
C_\varrho(\f{k},t)
=
\int d^3r\,\exp\{i\f{k}\cdot\f{r}\}
\langle\delta\hat\varrho_-({\bm 0},t)\delta\hat\varrho_-(\f{r},t)\rangle
\,. 
\ea
Note that the expectation values in Eqs.~(\ref{phase-phase}) and 
(\ref{density-density}) are realized via statistical averages in any 
measurement since the quantum state is not an eigenstate.  

To calculate the density fluctuations, we have to derive the 
time-dependence of the phase fluctuations beyond the zeroth-order
(frozen) part $\phi_{\fk{k}}^0$.
The approximate solution of Eq.\,(\ref{damping}) respectively 
Eq.\,(\ref{conformaldamping}) for frozen inflaton 
modes, transforming back to laboratory time, reads
\bea
\phi_{\fk{k}}(t)=\phi_{\fk{k}}^0
\left(1+\frac{c_0^2k^2}{2H^4t^2}\right)
+\ord\left(\frac{1}{t^3}\right) .
\ea
Since $g\propto1/t^4$, the  density fluctuations increase 
linearly in time, cf.~Eq.~(\ref{Bernoulli}): 
\bea
\delta\varrho_{\fk{k}}=-\frac{\dot\phi_{\fk{k}}}{g}=
\phi_{\fk{k}}^0\frac{c_0^2k^2}{g_0}\,t
+\ord\left(t^0\right). \label{rhofluct}
\ea
The overdot denotes 
a partial derivative with respect to lab time.
The sub-leading term $\ord\left(t^0\right)$ ensures the validity of the 
canonical 
commutation relation for hydrodynamical density and phase
operators, $[\delta\hat\varrho(t,{\bm r}),\hat\phi(t,{\bm r}')]
= i\delta ({\bm r}-{\bm r}')$.

Using the above relations (\ref{rhofluct}) and (\ref{phasefluct}), 
and inserting into  (\ref{density-density})
yields 
\bea
C_\varrho(\f{k},t) = \frac{H^2 c_0 k}{2g_0} t^2
\,.
\label{estC}
\ea
In addition to the assumptions already discussed in the case of 
phase-phase correlations ($c_0k \gg H \gg c_{\rm final}k$ and  
$k\xi_-(t) \ll1$), the total time duration of the sweep is limited by the 
maximally possible variation of the coupling 
$t_{\rm final}^2 = c_0/(c_{\rm final}H^2)$,
 according to Eq.~(\ref{c(t)}).
Again we estimate the maximally obtainable effect by setting 
$H \sim c_0k$, $k \sim k_{\rm final}^{\rm Planck}$,
and $t_{\rm final}^2 = c_0/(c_{\rm final}H^2)$.
Inserting into Eq.~(\ref{estC}), the maximally
obtainable relative change induced in 
the two-point density-density correlation function in the wave-number 
interval $\Delta k$ reads 
\bea
\frac{(\Delta k)^3C_\varrho^{\rm max}(\f{k},t_{\rm final})}{\varrho_0^2}
& \sim & 
\frac{H}{c_{\rm final}c_0}\,\frac{(\Delta k)^3}{\varrho_0} 
\nonumber\\
& \sim & 
\sqrt{\varrho_0 g_{\rm final}^3} \, 
\left(\frac{\Delta k}{ k_{\rm final}^{\rm Planck} }\right)^3
\,, 
\label{maxCorrrho}
\ea
where the last relation also uses, from 
$k_{\rm final}^{\rm Planck} \sim \sqrt{2\varrho_0 g_{\rm final}}$, 
that  $H\approx c_{\rm final} \sqrt{\varrho_0 g_0}= c_{\rm final} c_0 $.

The density-density correlations are increasing with time.  
However, the maximal relative change is still by a factor of 
$\sqrt{g_0/g_{\rm final}}\gg1$ smaller than in the case of the 
phase-phase correlations, Eq.\,(\ref{maxCorrphi}).
For density-density as well as phase-phase quantum correlations, 
the final outcome is rather small owing to the 
intrinsic smallness of the final diluteness parameter 
$\sqrt{\varrho_0 g_{\rm final}^3}$.

%%%%%%%%%%%%%%%%%%%%%%%%%%%%%%%%%%%%%%%%%%%%%%%%%%%%%%%%%%%%%%%%%%%%%%%%%%%%%%%
\subsection{Amplification in unstable regime}
%%%%%%%%%%%%%%%%%%%%%%%%%%%%%%%%%%%%%%%%%%%%%%%%%%%%%%%%%%%%%%%%%%%%%%%%%%%%%%%

It became evident in the preceding considerations that the impact of the 
frozen vacuum fluctuations is in principle observable, but rather weak.
Therefore, it is desirable to find some mechanism for the amplification 
of the fluctuations before measuring them.
As one possibility, we propose changing the sign of $g_-$ after the 
inflation phase, i.e., switching to the unstable regime,
for a short period of time.
If this change in $g_-$ occurs much faster than the (frozen) dynamics of
the field $\phi_-$, we may apply the sudden approximation, assuming
a step-function like change of $g_-$ from a small and positive 
$g^{\rm in}_-$ to a small negative value $g^{\rm out}_-$.

Within the sudden approximation, the relation of the behavior of the 
field $\phi_-$ just before ($\phi_-^{\rm in}$, $\dot\phi_-^{\rm in}$) 
and after ($\phi_-^{\rm out}$, $\dot\phi_-^{\rm out}$) the rapid change 
of $g_-$ from $g_-^{\rm in}>0$ to $g_-^{\rm out}<0$ can be obtained from 
the equations of motion
\bea
\phi_-^{\rm in}(\f{r},t) &=& \phi_-^{\rm out}(\f{r},t)
\,,
\nn
\frac{1}{g_-^{\rm in}}\,\dot\phi_-^{\rm in}(\f{r},t) 
&=& 
\frac{1}{g_-^{\rm out}}\,\dot\phi_-^{\rm out}(\f{r},t)
\,.
\ea
The latter equality is equivalent to  
$\delta\varrho_-^{\rm in}(\f{r},t)=\delta\varrho_-^{\rm out}(\f{r},t)$.
Assuming a constant $g^{\rm out}_-<0$ after the transition to the
unstable region, the subsequent time evolution is simply given by the 
superposition of the two independent solutions of imaginary spin mode 
frequency  
\bea
\phi^{\rm out}_{\fk{k}}(t)
&=&
A_{\fk{k}}\,
\exp\left\{+\sqrt{\frac{\varrho_0|g^{\rm out}_-|\f{k}^2}{m}}\,t \right\}
\nn
& &+
B_{\fk{k}}\,
\exp\left\{-\sqrt{\frac{\varrho_0|g^{\rm out}_-|\f{k}^2}{m}}\,t \right\}
\,.
\ea
Fluctuations with large wavenumbers 
(but still small compared to the  Planck scale) grow faster.
Note that this behavior is opposite to that of the gravitational (Jeans) 
instability in the early universe, where small wave-numbers experience a
stronger amplification -- larger structures collapse faster due
to the gravitational attraction than inhomogeneities on smaller length 
scales \cite{peebles,GravityinBEC}. 

If we had $\dot\phi^{\rm out}_{\fk{k}}/\phi^{\rm out}_{\fk{k}}=
-\sqrt{\varrho_0|g^{\rm out}_-|\f{k}^2/m}$, 
the factor $A_{\fk{k}}$ and hence the growing component of the 
solution would exactly vanish.
However, during the phase of freezing, the time-derivative 
$\dot\phi_{\fk{k}}$ decreases and is finally suppressed by a factor of 
$\sqrt{c_{\rm final}/c_0}$.
Hence the exponentially increasing part $A_{\fk{k}}$ does contribute 
in a roughly equal amount, $A_{\fk{k}} \approx B_{\fk{k}}$.
As a result, we obtain a drastic amplification of the phase 
(and also density) fluctuations until the non-linear regime 
(that is, phase separation \cite{Timmermans}) is reached, where the
fluctuations are not 
small compared to the order parameter itself anymore.
Consequently, one would expect the fluctuations to be measurable -- 
with state-selective absorption imaging, for example -- provided that
all other perturbations are small enough (see the next section).

%%%%%%%%%%%%%%%%%%%%%%%%%%%%%%%%%%%%%%%%%%%%%%%%%%%%%%%%%%%%%%%%%%%%%%%%%%%%%%%
%%%%%%%%%%%%%%%%%%%%%%%%%%%%%%%%%%%%%%%%%%%%%%%%%%%%%%%%%%%%%%%%%%%%%%%%%%%%%%%
\section{Conclusions}\label{Conclusions}
%%%%%%%%%%%%%%%%%%%%%%%%%%%%%%%%%%%%%%%%%%%%%%%%%%%%%%%%%%%%%%%%%%%%%%%%%%%%%%%
%%%%%%%%%%%%%%%%%%%%%%%%%%%%%%%%%%%%%%%%%%%%%%%%%%%%%%%%%%%%%%%%%%%%%%%%%%%%%%%

In summary, it is possible to simulate the behavior of the inflaton
field -- more accurately, its quantum fluctuations -- within
two-component Bose-Einstein condensates.
If we decrease $g_-$ with time according to Eq.~(\ref{c(t)}), the
phase fluctuations $\phi_-$ (i.e., the spin mode) experience an
effective de Sitter metric, and hence undergo the three stages of free 
oscillation, horizon crossing, and freezing explained in section 
\ref{Inflaton}.
The frozen initial quantum fluctuations can be measured after
an amplification phase, in which the coupling matrix is tuned to $g_-<0$.

The difference of the spectra (in $k$-space) of the frozen initial
quantum fluctuations (with an approximately scale invariant spectrum 
$\propto 1/k^3$)
and the usual quantum fluctuations of a condensate at rest 
($\propto 1/k$), provides one possibility to distinguish the signal from
other effects. 
As another option for a consistency check, one could keep the coupling  
$g_->0$ constant for a given time duration after the de Sitter phase 
(which has a decreasing $g_- \propto t^{-4} >0$) 
and before switching to the unstable regime $g_-<0$.
This way, one would allow the modes to freely oscillate again before
amplification, and thereby suppress those modes which are in the second
or fourth quarter of their period (i.e., have a decreasing amplitude)
at the moment of switching to the unstable regime.
This intermediate phase of small constant $g_- > 0$
would be roughly analogous to the epoch after
inflation, and the suppression mechanism is similar to the suppression of
modes in the cosmic microwave background, leading to the peaks and
valleys in its power spectrum \cite{microbackground}.  

%%%%%%%%%%%%%%%%%%%%%%%%%%%%%%%%%%%%%%%%%%%%%%%%%%%%%%%%%%%%%%%%%%%%%%%%%%%%%%
\subsection{Advantages}
%%%%%%%%%%%%%%%%%%%%%%%%%%%%%%%%%%%%%%%%%%%%%%%%%%%%%%%%%%%%%%%%%%%%%%%%%%%%%%

In comparison with the one-component case, the realization of an
effective geometry simulating the expansion of the universe
in a two-component Bose-Einstein condensate has several
advantages:
Firstly, in order to generate large relative variations of the spin 
coupling $g_-$, it is not necessary to go near the Feshbach resonance, 
and therefore the problems associated with molecule formation 
(inducing an imaginary part in the coupling constant) mentioned in the
introduction are not relevant.
Secondly, it is possible to change $g_-$ while sustaining a constant
background density:  
Assuming $V_a=V_b$ and $g_+=\rm const$, we obtain $\vau_a^0=0$ and 
$\varrho_-^0=0$, $\dot\varrho_+^0=0$ as a valid solution even for
time-dependent $g_-(t)$.
Thirdly, it is possible to amplify the fluctuations in the unstable
regime $g_-$ without necessarily destroying the condensate 
by phase separation.

%%%%%%%%%%%%%%%%%%%%%%%%%%%%%%%%%%%%%%%%%%%%%%%%%%%%%%%%%%%%%%%%%%%%%%%%%%%%%%
\subsection{Experimental parameters}
%%%%%%%%%%%%%%%%%%%%%%%%%%%%%%%%%%%%%%%%%%%%%%%%%%%%%%%%%%%%%%%%%%%%%%%%%%%%%%

In order to discuss the experimental feasibility of the proposed
quantum simulation of the inflaton, it is necessary to provide an
estimate for the experimental parameters involved.
If we assume that we may achieve essentially equal initial healing
lengths for the spin and the density mode  
$\xi^{\rm in}_-\approx \xi^{\rm in}_+=\ord(100\,{\rm nm})$, 
we may simulate an expansion of the universe corresponding to 
more than one $e$-folding (i.e., $H\Delta\tau>1$), arriving at the
final healing lengths  
$\xi^{\rm out}_-=\ord (1\,\mu\rm m)$ and
$\xi^{\rm out}_+=\xi^{\rm in}_+=\ord(100\,\rm nm)$.
Although one $e$-folding is tiny compared to the expansion
during inflation (which involves many orders of magnitude), one can
still reproduce generic features (such as freezing). 
Since the concept of the effective geometry only applies for
wavelengths large compared to the healing length  
(the analogue of the Planck scale), the characteristic size of the
interesting structures (and hence the size of the condensate cloud)
should be several micrometers -- which can be resolved with optical
methods (e.g., state-selective resonant absorption imaging). 

Furthermore, the measurability of the quantum fluctuations requires the
absence of any other fluctuations (noise) which may swamp the signal.
Assuming an initial speed of sound of about 1 mm/sec, a mode with a
wavelength of a few $\mu$m corresponds to an initial
frequency somewhere below 1 kHz.
Hence one should avoid temperatures above 100 nK, for which the
thermal fluctuations exceed the (to be measured) quantum fluctuations 
(assuming no thermalization during the experiment). 
For the parameters specified above, the Hubble parameter determining
the timescale for changing $g_-$ would also be below 1 kHz.
This should represent no problem as it is possible to change the
coupling constants (i.e., the external magnetic field) on the
time-scale of some microseconds. 

Although quite challenging, 
the proposed experiment should thus in principle 
be feasible with present-day technology.

%%%%%%%%%%%%%%%%%%%%%%%%%%%%%%%%%%%%%%%%%%%%%%%%%%%%%%%%%%%%%%%%%%%%%%%%%%%%%%
\subsection{Planckian problem}
%%%%%%%%%%%%%%%%%%%%%%%%%%%%%%%%%%%%%%%%%%%%%%%%%%%%%%%%%%%%%%%%%%%%%%%%%%%%%%

Apart from simulating the inflaton field experimentally 
(quantum simulation via analogues), one of the long-term motivations 
of the present work was to be able to ultimately 
investigate the impact of the behavior at the effective Planck scale
on the frozen fluctuations (as explained in the introduction).
For dilute Bose-Einstein condensates, the behavior for small
wavelengths (i.e., below the healing length) is in some sense trivial
-- the excitations are free particles of mass $m$ (the constituent atoms),  
with $\omega=\f{k}^2/(2m)$. 
More importantly, though, the healing length increases with time according to 
Eqs.~(\ref{healing}) and (\ref{c(t)}) -- i.e., it grows during the
expansion -- which is (almost) certainly not a realistic feature of the
real Planck length.
In order to circumvent this obstacle, one could introduce another
cut-off scale which is sensitive to frequencies instead of
wavenumbers.
For example, via exposing the condensate to monochromatic radiation
with a detuned radio-frequency $\omega=\omega_0-\Delta\omega$ with
$\omega_0$ being the frequency of a suitable hyperfine transition and
$\Delta\omega$ the detuning, one may couple frequency-selectively 
couple phonons with the frequency $\Delta\omega$ to that transition. 
In that case, $\Delta\omega$ would be a reasonable analogue of the Planck
scale, by fixing a frequency scale at which free phonon propagation 
is cut off.

\subsection{Outlook} 
After succeeding to measure the main effect -- the fact that 
quantum fluctuations of the inflaton field in de Sitter space-time
get frozen at a specific value -- we can start to 
manipulate the behavior at the effective Planck scale and study 
its impact on the frozen fluctuations. The exciting prospect of
undertaking such manipulations of Planck scale physics is that 
one were able to experimentally investigate analogue signatures of new 
(trans-Planckian) physics in the anisotropies of the cosmic microwave 
background; such possible signatures of trans-Planckian physics 
are  discussed, e.g., in \cite{signatures}.

We, finally, mention that a further interesting aspect 
of the scale separation of independently
propagating spin and density modes is the possible implementation 
of a variant of the supersonical ``warp-drives'' proposed 
in \cite{warp}. The reference frame, against which 
the ``superluminal'' motion of the $\phi_-$ mode is measured, 
is not provided by the laboratory frame, like in \cite{warp}, 
but by the $\phi_+$ mode, which has effectively instantaneous 
signal transfer on the time scales of the $\phi_-$ mode.
In addition, the ``hard'' $\phi_+$ mode is essentially flat on the
curvature radius scale of the $\phi_-$ mode, i.e., appears to be
 essentially Minkowski space-time as seen from the $\phi_-$ mode. 
The Minkowski background of the $\phi_+$ mode represents 
a necessary prerequisite for ``superluminal'' travel in warped space-times to 
be operational and definable, 
because it enables the comparison of two metrics, one of them
flat and the other curved, on the same manifold \cite{Olum}.
In contrast to the single Euler-fluid case of \cite{warp}, 
in the presently discussed two-component Bose-Einstein-condensed
gas both of these metrics obey Lorentz invariance. 

%%%%%%%%%%%%%%%%%%%%%%%%%%%%%%%%%%%%%%%%%%%%%%%%%%%%%%%%%%%%%%%%%%%%%%%%%%%%%%%
%%%%%%%%%%%%%%%%%%%%%%%%%%%%%%%%%%%%%%%%%%%%%%%%%%%%%%%%%%%%%%%%%%%%%%%%%%%%%%%
\acknowledgments
%%%%%%%%%%%%%%%%%%%%%%%%%%%%%%%%%%%%%%%%%%%%%%%%%%%%%%%%%%%%%%%%%%%%%%%%%%%%%%%
%%%%%%%%%%%%%%%%%%%%%%%%%%%%%%%%%%%%%%%%%%%%%%%%%%%%%%%%%%%%%%%%%%%%%%%%%%%%%%%

The authors thank J.\,I. Cirac, A.\,J.~Leggett, W.\,G.~Unruh,
G.\,E. Volovik, and M.~Weitz for fruitful discussions. 
R.\,S.~gratefully acknowledges financial support by the Emmy Noether
Programme of the German Research Foundation (DFG) under grant 
No.~SCHU 1557/1-1 and by the Humboldt foundation. 
U.\,R.\,F. and R.\,S. both acknowledge support by the COSLAB programme
of the ESF.  
  
%%%%%%%%%%%%%%%%%%%%%%%%%%%%%%%%%%%%%%%%%%%%%%%%%%%%%%%%%%%%%%%%%%%%%%%%%%%%%%%
%%%%%%%%%%%%%%%%%%%%%%%%%%%%%%%%%%%%%%%%%%%%%%%%%%%%%%%%%%%%%%%%%%%%%%%%%%%%%%%
%%%%%%%%%%%%%%%%%%%%%%%%%%%%%%%%%%%%%%%%%%%%%%%%%%%%%%%%%%%%%%%%%%%%%%%%%%%%%%%
%%%%%%%%%%%%%%%%%%%%%%%%%%%%%%%%%%%%%%%%%%%%%%%%%%%%%%%%%%%%%%%%%%%%%%%%%%%%%%%

%%%%%%%%%%%%%%%%%%%%%%%%%%%%%%%%%%%%%%%%%%%%%%%%%%%%%%%%%%%%%%%%%%%%%%%%%%%%%%%
%%%%%%%%%%%%%%%%%%%%%%%%%%%%%%%%%%%%%%%%%%%%%%%%%%%%%%%%%%%%%%%%%%%%%%%%%%%%%%%
%%%%%%%%%%%%%%%%%%%%%%%%%%%%%%%%%%%%%%%%%%%%%%%%%%%%%%%%%%%%%%%%%%%%%%%%%%%%%%%
%%%%%%%%%%%%%%%%%%%%%%%%%%%%%%%%%%%%%%%%%%%%%%%%%%%%%%%%%%%%%%%%%%%%%%%%%%%%%%%
%%%%%%%%%%%%%%%%%%%%%%%%%%%%%%%%%%%%%%%%%%%%%%%%%%%%%%%%%%%%%%%%%%%%%%%%%%%%%%%
%%%%%%%%%%%%%%%%%%%%%%%%%%%%%%%%%%%%%%%%%%%%%%%%%%%%%%%%%%%%%%%%%%%%%%%%%%%%%%%
%%%%%%%%%%%%%%%%%%%%%%%%%%%%%%%%%%%%%%%%%%%%%%%%%%%%%%%%%%%%%%%%%%%%%%%%%%%%%%%
%%%%%%%%%%%%%%%%%%%%%%%%%%%%%%%%%%%%%%%%%%%%%%%%%%%%%%%%%%%%%%%%%%%%%%%%%%%%%%%
%%%%%%%%%%%%%%%%%%%%%%%%%%%%%%%%%%%%%%%%%%%%%%%%%%%%%%%%%%%%%%%%%%%%%%%%%%%%%%%
%%%%%%%%%%%%%%%%%%%%%%%%%%%%%%%%%%%%%%%%%%%%%%%%%%%%%%%%%%%%%%%%%%%%%%%%%%%%%%%
%%%%%%%%%%%%%%%%%%%%%%%%%%%%%%%%%%%%%%%%%%%%%%%%%%%%%%%%%%%%%%%%%%%%%%%%%%%%%%%
%%%%%%%%%%%%%%%%%%%%%%%%%%%%%%%%%%%%%%%%%%%%%%%%%%%%%%%%%%%%%%%%%%%%%%%%%%%%%%%
%%%%%%%%%%%%%%%%%%%%%%%%%%%%%%%%%%%%%%%%%%%%%%%%%%%%%%%%%%%%%%%%%%%%%%%%%%%%%%%
%%%%%%%%%%%%%%%%%%%%%%%%%%%%%%%%%%%%%%%%%%%%%%%%%%%%%%%%%%%%%%%%%%%%%%%%%%%%%%%
%%%%%%%%%%%%%%%%%%%%%%%%%%%%%%%%%%%%%%%%%%%%%%%%%%%%%%%%%%%%%%%%%%%%%%%%%%%%%%%
%%%%%%%%%%%%%%%%%%%%%%%%%%%%%%%%%%%%%%%%%%%%%%%%%%%%%%%%%%%%%%%%%%%%%%%%%%%%%%%

\begin{thebibliography}{499}
%%%%%%%%%%%%%%%%%%%%%%%%%%%%%%%%%%%%%%%%%%%%%%%%%%%%%%%%%%%%%%%%%%%%%%%%%%%%%%%
%%%%%%%%%%%%%%%%%%%%%%%%%%%%%%%%%%%%%%%%%%%%%%%%%%%%%%%%%%%%%%%%%%%%%%%%%%%%%%%
%%%%%%%%%%%%%%%%%%%%%%%%%%%%%%%%%%%%%%%%%%%%%%%%%%%%%%%%%%%%%%%%%%%%%%%%%%%%%%%
%%%%%%%%%%%%%%%%%%%%%%%%%%%%%%%%%%%%%%%%%%%%%%%%%%%%%%%%%%%%%%%%%%%%%%%%%%%%%%%

\bibitem{Linde} A.\,D. Linde, 
%{\em Chaotic inflation}
Phys. Lett. B {\bf 129}, 177 (1983). 

\bibitem{Lidsey} J.\,E. Lidsey, A.\,R. Liddle, E.\,W. Kolb, E.\,J. 
Copeland, T. Barreiro, and M. Abney, 
%: {\em Reconstructing the inflaton potential--an overview}, 
Rev. Mod. Phys. {\bf 69}, 373 (1997).  

\bibitem{inflation} A.\,H. Guth, Phys. Rev. D \textbf{23}, 347 (1981).

\bibitem{GrishaBook} G.\,E. Volovik, 
\textit{The Universe in a Helium Droplet} 
(Oxford University Press, Oxford, 2003). 

\bibitem{unruh} W.\,G. Unruh, 
%: {\em Experimental Black-Hole Evaporation?}, 
Phys. Rev. Lett. {\bf 46}, 1351 (1981); 
Phys. Rev. D {\bf 51}, 2827 (1995).

\bibitem{MattPRL} M. Visser, 
%: {\em Hawking radiation without black hole entropy}, 
Phys. Rev. Lett. {\bf 80}, 3436 (1998).


\bibitem{AG-BEC} 
L.\,J.\,Garay, J.\,R.\,Anglin, J.\,I.\,Cirac, and P.\,Zoller,
%{\em Sonic Analog of Gravitational Black Holes in Bose-Einstein Condensates},
Phys.\ Rev.\ Lett.\ {\bf 85}, 4643 (2000);
%
%L.\,J.\,Garay, J.\,R.\,Anglin, J.\,I.\,Cirac, and P.\,Zoller,
%{\em Sonic black holes in dilute Bose-Einstein condensates},
Phys.\ Rev.\ A {\bf 63}, 023611 (2001);
%
L.\,J.\,Garay,
%{\em Black Holes In Bose-Einstein Condensates},
Int.\ J.\ Theor.\ Phys.\  {\bf 41}, 2073 (2002).

\bibitem{BHdispers} T. Jacobson, Phys. Rev. D {\bf 53}, 7082 (1996); 
S. Corley and T. Jacobson, 
%: {\em Hawking spectrum and high frequency dispersion}, 
Phys. Rev. D {\bf 54}, 1568 (1996). 

\bibitem{remark} 
We remark that, in our present analogue gravity context, the analogue
inflaton field emerges naturally (as a phonon mode), and is not
introduced {\it ad hoc} into the theory.
%
%We remark that in our present analogue gravity 
%context, the inflaton is itself a matter field (of phonons), and 
%not an extraneous scalar field introduced {\it ad hoc}.
% into the theory. 

\bibitem{BLVBEC} C.\,Barcel\'o, S.\,Liberati, and M.\,Visser,
%{\em Analogue gravity from Bose-Einstein condensates},
Class.\ Quantum\ Grav.\  {\bf 18}, 1137 (2001).

\bibitem{uwe+petr} P.\,O. Fedichev and U.\,R. Fischer, 
Phys. Rev. Lett. {\bf 91}, 240407 (2003); 
Phys. Rev. D {\bf 69}, 064021 (2004); %.\bibitem{MPLA} 
U.\,R. Fischer, Mod. Phys. Lett. A {\bf 19}, 1789 (2004).
%{\tt cond-mat/0406086}.

\bibitem{0303063} P.\,O. Fedichev and U.\,R. Fischer, 
Phys. Rev. A {\bf 69}, 033602  (2004).

\bibitem{BLVPRA} C.\,Barcel\'o, S.\,Liberati, and M.\,Visser, 
Phys. Rev. A {\bf 68}, 053613 (2003).

\bibitem{Piotr} P.\,O. Fedichev, M.\,W. Reynolds, and G.\,V. Shlyapnikov, 
%: {\em Three-Body Recombination of Ultracold Atoms to 
%a Weakly Bound s Level}, 
Phys. Rev. Lett. {\bf 77}, 2921 (1996).

\bibitem{Weber} T. Weber, J. Herbig, M. Mark, H.-C. 
N\"agerl, and R. Grimm,  
Phys. Rev. Lett. {\bf 91}, 123201 (2003). 


\bibitem{HoShenoy} 
T.-L. Ho and V.\,B.\,Shenoy, 
Phys.\ Rev.\ Lett.\ {\bf 77}, 3276 (1996).

\bibitem{Timmermans} E. Timmermans, Phys. Rev. Lett. {\bf 81}, 5718 (1998).

\bibitem{Hall} 
D.\,S.\,Hall, M.\,R.\,Matthews, J.\,R.\,Ensher, C.\,E.\,Wieman, and
E.\,A.\,Cornell,  
Phys.\ Rev.\ Lett.\ {\bf 81}, 1539 (1998); 
D.\,S.\,Hall, M.\,R.\,Matthews, C.\,E.\,Wieman, and E.\,A.\,Cornell, 
{\it ibid.} {\bf 81}, 1543 (1998).

\bibitem{Miesner} 
H.-J.\,Miesner, D.\,M.\,Stamper-Kurn, J.\,Stenger, S.\,Inouye, 
A.\,P.\,Chikkatur, and W.\,Ketterle, 
Phys.\ Rev.\ Lett.\ {\bf 82}, 2228 (1999).

\bibitem{CalzettaHu} E.\,A. Calzetta and B.\,L. Hu, 
Phys. Rev. A {\bf 68}, 043625 (2003). 


\bibitem{PGL} P.\,Painlev{\'e}, 
%{\em La M\'ecanique classique et la th\'eorie de la relativit\'e},
C.\ R.\ Hebd.\ Seances Acad.\ Sci.\ (Paris) {\bf 173}, 677 (1921);
%
A.\,Gullstrand, 
%{\em Allgemeine L\"osung des statischen Eink\"orper\-problems in der
%Einsteinschen Gravitations\-theorie},
Ark.\ Mat.\ Astron.\ Fys.\ {\bf 16}, 1 (1922);
%
G.\,Lema{\^\i}tre, 
%{\em L'univers en expansion},
Ann.\ Soc.\ Sci.\ (Bruxelles) A {\bf 53}, 51 (1933).

\bibitem{BLV2002} C.\,Barcel\'o, S.\,Liberati, and M.\,Visser,
Class.\ Quantum\ Grav.\  {\bf 19}, 2961 (2002).

\bibitem{slow-light}
W.\,G.~Unruh and R.~Sch\"utzhold,
%``On slow light as a black hole analogue,''
Phys.\ Rev.\ D {\bf 68}, 024008 (2003).
%[arXiv:gr-qc/0303028].
%%CITATION = GR-QC 0303028;%%

\bibitem{dbha}
R.~Sch\"utzhold, G.~Plunien, and G.~Soff,
%``Dielectric black hole analogues,''
Phys.\ Rev.\ Lett.\  {\bf 88}, 061101 (2002).
%[arXiv:quant-ph/0104121].
%%CITATION = QUANT-PH 0104121;%%

\bibitem{visser} 
M.\,Visser, 
%: {\em Acoustic black holes: horizons, ergospheres, and Hawking radiation}, 
Class.\ Quantum Grav.\ {\bf 15}, 1767 (1998).

\bibitem{birrell}
N.\,D.~Birrell and P.\,C.\,W.~Davies,
{\em Quantum Fields in Curved Space},
(Cambridge University Press, Cambridge, England, 1982).

\bibitem{peebles}
P.\,J.\,E.~Peebles, 
{\em Principles of Physical Cosmology}
(Princeton University Press, Princeton, 1993). 

 \bibitem{GravityinBEC} Note, however, that an attractive 
$1/r$ ``gravity''-interaction between the atoms may 
be induced by shining off-resonant intense laser beams onto a BEC:  
D. O'Dell, S. Giovanazzi, G. Kurizki, and V.\,M. Akulin, 
%: {\em Bose-Einstein Condensates with $1/r$ Interatomic Attraction:
% Electromagnetically Induced ``Gravity''}, 
Phys. Rev. Lett. {\bf 84}, 5687 (2000). 
%
Such a deviation from the usual dilute-gas contact interaction 
can in principle give rise to a (classical) gravitational ``clumping'' 
tendency which is analogous to the Jeans instability.

\bibitem{tof}
W.~Ketterle, D.\,S.~Durfee, and D.\,M.~Stamper-Kurn,
{\em Making, probing and understanding Bose-Einstein condensates}, 
in Proc.\ 1998 Fermi Summer School on BEC in Varenna/Italy 
%in: Proceedings of the 1998 Enrico Fermi summer school
%on Bose-Einstein condensation in Varenna, Italy;
(cond-mat/9904034).

\bibitem{microbackground} P. de Bernardis {\it et al.},
%: {\em A flat Universe from high-resolution maps of the 
%cosmic microwave background radiation}, 
Nature {\bf 404}, 955 (2000). 

\bibitem{signatures}  A. Kempf,
%``Mode generating mechanism in inflation with cutoff,''
Phys. Rev. D {\bf 63}, 083514 (2001); 
J.\,C.~Niemeyer and R.~Parentani, 
%``Trans-Planckian dispersion and scale-invariance of inflationary
%perturbations,''
Phys.\ Rev.\ D {\bf 64}, 101301(R) (2001);
J.~Martin and R.\,H.~Brandenberger,
%``The Corley-Jacobson dispersion relation and trans-Planckian inflation,''
Phys.\ Rev.\ D {\bf 65}, 103514 (2002).
%; A.~Kempf and J.\,C.~Niemeyer,
%``Perturbation spectrum in inflation with cutoff,''
%Phys.\ Rev.\ D {\bf 64}, 103501 (2001); 
%J.~Martin and R.~Brandenberger,
%``On the dependence of the spectra of fluctuations in inflationary  cosmology
%on trans-Planckian physics,''
%Phys.\ Rev.\ D {\bf 68}, 063513 (2003).
%U.~H.~Danielsson,
%``A note on inflation and transplanckian physics,''
%Phys.\ Rev.\ D {\bf 66} (2002) 023511;
%R.~Easther, B.~R.~Greene, W.~H.~Kinney and G.~Shiu,
%``A generic estimate of trans-Planckian modifications to the primordial power
%spectrum in inflation,''
%Phys.\ Rev.\ D {\bf 66} (2002) 023518.
%N.~Kaloper, M.~Kleban, A.~E.~Lawrence and S.~Shenker,
%``Signatures of short distance physics in the cosmic microwave  background,''
%Phys.\ Rev.\ D {\bf 66} (2002) 123510.

\bibitem{warp} U.\,R.\,Fischer and M.\,Visser,   
%{\em Warped space-time for phonons moving in a 
%perfect nonrelativistic fluid}, 
%gr-qc/0211029,  
Europhys.\ Lett.\ {\bf 62}, 1 (2003).

\bibitem{Olum} 
K.\,D. Olum, 
%: {\em Superluminal Travel Requires Negative Energies}, 
Phys. Rev. Lett. {\bf 81}, 3567 (1998).

\end{thebibliography}
\end{document}